\documentclass[]{article}

\usepackage{arxiv}

\usepackage[utf8]{inputenc} 
\usepackage[T1]{fontenc}    
\usepackage{hyperref}       
\usepackage{url}            
\usepackage{booktabs}       
\usepackage{amsfonts}       
\usepackage{amsmath}
\usepackage{nicefrac}       
\usepackage{microtype}      
\usepackage{cleveref}       
\usepackage{graphicx}
\usepackage{doi}

\usepackage{cite}
\usepackage{subfig}
\usepackage{tabularx}
\usepackage[ruled,vlined,linesnumbered]{algorithm2e}
\usepackage{import} 
\usepackage[]{siunitx}
\usepackage{csquotes} 
\usepackage{multirow}
\usepackage{array}
\usepackage{orcidlink}

\DeclareMathOperator*{\argmax}{arg\,max}

\title{Efficient Conflict Graph Creation for Time-Sensitive Networks with Dynamically Changing Communication Demands}

\author{\orcidlink{0000-0002-5146-0184}Heiko~Geppert\\
	IPVS\\
	University of Stuttgart\\
	Germany\\
	\texttt{heiko.geppert@uni-stuttgart.de} \\
	\And
	\orcidlink{0000-0002-3470-7712}Frank~D\"urr\\
	IPVS\\
	University of Stuttgart\\
	Germany\\
	\texttt{frank.duerr@uni-stuttgart.de} \\
	\And
	\orcidlink{0000-0001-8986-8241}Kurt~Rothermel\\
	IPVS\\
	University of Stuttgart\\
	Germany\\
	\texttt{kurt.rothermel@uni-stuttgart.de} \\
}

\renewcommand{\shorttitle}{Efficient Conflict Graph Creation}

\hypersetup{ 
pdftitle={Efficient Conflict Graph Creation for Time-Sensitive Networks with Dynamically Changing Communication Demands},
pdfsubject={cs.NI},
pdfauthor={Heiko Geppert et al.},
pdfkeywords={deterministic networks, stream scheduling, TSN, conflict graph, scalability},
}

\begin{document}
\maketitle

\begin{abstract}
	Many applications of cyber-physical systems require real-time communication: manufacturing, automotive, etc.
	Recent Ethernet standards for Time Sensitive Networking (TSN) offer time-triggered scheduling in order to guarantee low latency and jitter bounds.
	This requires precise frame transmission planning, which becomes especially hard when dealing with many streams, large networks, and dynamically changing communications.
	A very promising approach uses conflict graphs, modeling conflicting transmission configurations.
	Since the creation of conflict graphs is the bottleneck in these approaches, we provide an improvement to the conflict graph creation.
	We present a randomized selection process that reduces the overall size of the graph in half and three heuristics to improve the scheduling success.
	In our evaluations we show substantial improvements in the graph creation speed and the scheduling success compared to existing work, updating existing schedules in fractions of a second.
	Additionally, offline planning of 9000 streams was performed successfully within minutes.
\end{abstract}

\keywords{deterministic networks \and stream scheduling \and TSN \and conflict graph \and scalability}


\section{Introduction}
\label{sec:introduction}

Real-time communication with deterministic bounds on network delay is an essential requirement for many cyber-physical systems (CPS) consisting of networked sensors, actuators, and controllers. 
Such CPS are often safety-critical. 
The late arrival of sensor values or actuator commands can have severe or even catastrophic consequences, such as damaged machinery in automated factories or storage facilities.
Other popular examples include smart energy grids with the danger of blackouts if real-time properties are violated.

The need for deterministic real-time communication technologies has been recognized by industry and standardization institutes alike. 
The  Institute of Electrical and Electronics Engineers (IEEE) has defined a set of mechanisms for real-time Ethernet, commonly known by the term Time-Sensitive Networking (TSN). 

In particular, the standard IEEE 802.1Qbv (Enhancements for Scheduled Traffic) \cite{IEEE802.1Qbv-2015} defines the Time-Aware Shaper (TAS), which allows for deterministic frame forwarding in the range of microseconds for time-triggered traffic (also known as scheduled traffic).
Using the TSN standards, time-triggered streams can be scheduled precisely even in networks with tens to hundreds of network elements.

A common NP-hard problem is to define schedules for the TAS to find spatially and temporally isolated frame transmission slots throughout the network, guaranteeing a timely delivery of every single time-triggered frame.
The computation of such traffic plans is not part of the standard.
Consequently, vendors and operators can create their own solutions tailored to their specific use case, which made the computation of traffic plans the object of many research endeavors \cite{Stueber2022}.
A common approach is to define constraints for an Integer Linear Program, Constraint Program, or Satisfiability Modulo Theories, and to compute a traffic plan with a generic solver.
However, this does not scale to large networks or dynamically changing systems, requiring fast incremental updates, leading to the investigation of heuristics.

A well-known approach to simplify the scheduling process is to use the no-wait principle \cite{Duerr2016}.
Frames arriving at a bridge are forwarded immediately without buffering.
This results in low latency and jitter, and leads to deterministic system behavior. 
Furthermore, the no-wait principle requires only one egress queue of the TAS, leaving the remaining queues free for other traffic types.
The gate control list (GCL) managing the gates in the Time-Aware Shaper is minimized, since the one gate used by the no-wait traffic is always opened.
Hence, no additional GCL entries are needed for the time-triggered traffic.
Using the no-wait approach reduces the solution space which can lead to worse results \cite{Vlk2022}, i.e., lower schedulability.
In practice, wait-allowed (queuing) severely reduces the scalability and can be counterproductive for the schedulability in resource-limited systems \cite{Xue2023}, e.g., when expecting short runtimes.

In 2020, Falk et al. proposed a conflict graph based no-wait scheduling approach \cite{Falk2020}.
The results from the original paper and an analysis in another work \cite{Xue2023} identify this as a very promising scheduling approach, worthwhile of further investigation and optimization.
The basic idea is to translate the constraints of the scheduling problem---in particular the spatial and temporal separation---into a conflict graph, and then solve an independent colorful set (ICS) problem to find a valid traffic plan.
Later the approach was extended, to cover dynamic scenarios \cite{Falk2022}.
Thereby, changes in the scheduled communication, e.g., adding new streams, is covered by updating the conflict graph and solving the ICS problem on the modified graph, while keeping QoS guarantees intact.
The process of creating conflict graphs and of updating them works identically.
For this reason, we refer to both as conflict graph expansion.

There are two major problems with the existing conflict graph expansion: 
First, configurations are added to conflict graphs agnostic to any additional information to guide this process. 
We call this uninformed process \emph{homogeneous} graph expansion, since all streams are treated equally with respect to the number of explored configurations.
Secondly, recurrent graph structures for the same configurations might be created deterministically and recurrently, even if they have failed to establish solutions previously.
This is especially inefficient, since the number of nodes in the conflict graph has a major impact on the time needed to find schedules and react to system changes, like adding additional streams.

Our contribution is an improvement of the conflict graph expansion.
We propose a new \emph{informed} and \emph{randomized} approach. 
Heuristic information is used to steer graph generation into directions that are more likely to find solutions. 
Thereby, we present three different heuristics, using either stream properties (the streams traffic volume) or conflict graph metrics (average degree or page rank in the conflict graph). 
Randomization is used to avoid creating recurrent graph structures that did not lead to solutions previously and explore new structures instead, which reduces the required conflict graph size to about one third.
This enables the conflict graph-based scheduling approach to be used in large scale environments while keeping the computation time and memory requirement manageable. 

The rest of the paper is structured as follows. 
In Section~\ref{sec:related-work}, we discuss the related work.
We present the system model in Section~\ref{sec:system-model}.
The scheduling problem, the conflict graph, and the independent colorful set problem, are introduced in Section~\ref{sec:preliminaries}.
The problem formalization of creating efficient conflict graphs is given in Section~\ref{sec:problem-statement}.
Our proposed approaches are shown in Section~\ref{sec:conflict_graphs} and evaluated in Section~\ref{sec:eval}.
Finally, we conclude the paper in Section~\ref{sec:conclusion}.


\section{Related Work}
\label{sec:related-work}

Time-triggered traffic planning in real-time networks has become an intensively studied field in recent years.
One popular way to compute traffic plans is to use mathematical solvers like Integer Linear Programming, Constraint Programming, or Satisfiability Modulo Theory \cite{Santos2019, Steiner2010, SernaOliver2018, Falk2018, Vlk2021}.
These approaches have the benefit of yielding optimal results for a given constrained optimization problem at the cost of high runtimes.

Especially in larger systems with many time-triggered streams and many bridges, the runtime can grow to several days \cite{Geppert2023}, which is infeasible for scheduling dynamic systems.
In consequence, heuristics have been developed that can compute traffic plans within reasonable time.
Vlk et al. proposed a backtracking-based heuristic \cite{Vlk2022}.
They use a conflict storage which serves a similar purpose as the conflict graph that we use.
Their approach covers the possible solution space by backtracking, leading to about one order of magnitude longer runtimes than ours.
A notion of flexibility to allow for future changes was defined by G\"{a}rtner et al. alongside a fast scheduling heuristic \cite{Gaertner2022}.
However, they only evaluate their heuristic in very small networks, without any meaningful routing options.
Raagaard et al. suggested a set of very fast heuristics with per stream assignment \cite{Raagaard2017}.
This is computationally efficient, at the cost of admitting fewer streams.
In contrast, our approach allows for batch updates and takes the whole batch into consideration throughput the assignment.
Further, their main objective is to minimize the number of required queues.
Our no-wait approaches inherently requires only a single queue.

The most promising heuristic for dynamic stream sets is the approach of Falk et al. \cite{Falk2022}, which employs conflict graphs.
Conflict graphs model violations of spatial and temporal isolation between scheduling options of time-triggered streams.
This allows to encode the configuration space of the traffic planning problem and to solve it using efficient graph algorithms like the Greedy Flow Heap Heuristic (GFH).
Thereby, it is sufficient to work with subgraphs, i.e., to find a traffic plan based on a subset of the whole configuration space.
This way, the space and time effort spent on the problem can be regulated.
Another feature to reduce the computation time used in this approach is the use of defensive and offensive planning.
Whenever new streams are to be added, they first use the old solution and try to add valid configurations for the new streams.
If this does not work they consider a complete recomputation and take the better solution.
However, the conflict graph creation procedure still imposes a substantial bottleneck, because too many nodes are added to the graph and each node requires conflict checks against all nodes already in the graph.
Our approach to build conflict graphs solves this problem by randomizing the phase of new configurations, so that even larger problem instances are solvable within a short time.
Additionally, we create configurations in an informed way to improve the scheduling success further.


\section{Network and Stream Model}
\label{sec:system-model}

In the following, we discuss our network model, stream properties, and scheduling assumptions.

In this work, we propose a solution for time-driven scheduling for periodical isochronous traffic \cite{IECSC65C/WG18}.
We assume a full-duplex \textbf{network} $\mathcal{N}$, where all bridges and end stations have synchronized clocks to enable precise frame scheduling operations, i.e., frames can be scheduled with \SI{1}{\micro\second} \emph{macro tick} granularity.

Our scheduling approach conforms to the time-sensitive networking standards.
Deterministic frame forwarding is provided by the Time Aware Shaper (TAS), which is specified in the IEEE 802.1 Qbv standard \cite{IEEE802.1Qbv-2015}.
Each bridge implements up to eight egress queues per egress port. 
Traffic is mapped to one of the egress queues through a dedicated header field in the Ethernet frame, the so-called Priority Code Point. 
Each egress queue is guarded by gates, which open and close based on a schedule (timetable) called the Gate Control List (GCL). 
Frames within a queue can be transmitted only if the gate is open. 
All bridges are synchronized through the gPTP protocol defined in IEEE 820.1AS \cite{IEEE802.1AS-2020} such that the GCLs at different bridges can control the timely forwarding of frames from talkers (sources) to listeners (destinations).
Further, we assume a store-and-forward model, meaning that each bridge must wait until the entire frame is received before forwarding it, and per-stream routing \cite{IEEE802.1Qca-2015}.

Scheduling and routing is orchestrated by a central network entity in accordance with IEEE 802.1 Qcc \cite{IEEE802.1Qcc-2018}, which creates the Centralized Network Configuration (CNC).
We assume that the scheduler has full knowledge about the network topology and the currently deployed traffic plan to handle upcoming scheduling requests.
Whenever the traffic plan needs to be updated, all update requests are handled in one batch at the central entity, which then computes candidate routes for the new streams and tries to find a valid assignment.

\label{sec:system-model:stream-definition}

Every \textbf{time-triggered stream} is defined as a tuple $s = ($\emph{src}, \emph{dst}, \emph{period}, \emph{size}, \emph{deadline}$)$, with \emph{src} and \emph{dst} denoting the stream's source and destination.
In our application model, streams always start and end at end devices, while bridges only forward frames.
The \emph{period} indicates the cycle time of the stream, and \emph{size} is the frame size in bytes.
We assume constant frame sizes, e.g., a sensor always transmitting frames with the same payload size.
Further, we expect the frame sizes to include all headers and inter-frame gaps, so that we can plan them back-to-back without interference from other network layers.
The \emph{deadline} is the latest time, relative to the start of the period, that a frame must arrive at \emph{dst}.
We assume $\emph{deadline} = \emph{period}$ for all streams, so that frames are released (ready to be sent) at the start of their period and need to be delivered within their period to prevent interference with later periods.
All properties in the tuple are static stream characteristics.

We use the no-wait principle \cite{Duerr2016} for all streams.
This means, every frame is forwarded as soon as it is received by a network device.
A side effect of the no-wait principle is, that we only need a single egress queue for our time-triggered streams, because they are not queued, i.e., transmitted as soon as they enter the egress queue.
With the no-wait principle, we also gain a better plannability since the arrival time of a frame solely depends on its route and initial phase.
Therefore, we define two variable properties: route $\pi$ (from a set of candidate routes) and phase $\phi$ (transmission time offset at \emph{src}).
In a network with no-wait principle and known propagation, transmission, and processing delays, we can always determine the position of any frame if we also have the stream characteristics, route, and phase.
Hence, precise traffic plans can be created by defining $\pi$ and $\phi$ for each stream.

Using a set of candidate routes provides a middle ground between fixed routing and integrating the routing within the scheduling. 
Note that a set size of one, i.e., fixed routing is possible.
Further, routing constraints can be delegated to the routing algorithm easily.
So, in case a stream needs to pass a certain device, e.g., for some event processing, this can be included in the routing and does not affect the scheduler's mode of operation.

\begin{figure}
	\centering
	\includegraphics[width=0.5\linewidth]{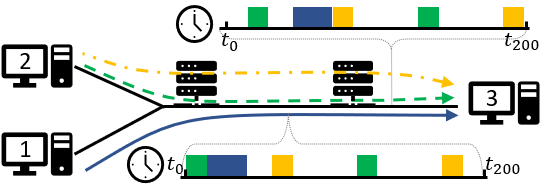}
	\caption{Example system. The blue stream has a frame size of \SI{500}{\byte} and a period of \SI{200}{\micro\second}. The green and yellow streams both have frame sizes of \SI{250}{\byte} and periods \SI{100}{\micro\second}.}
	\label{fig:systemmodelexample}
\end{figure}

We display an exemplary system in Figure~\ref{fig:systemmodelexample}.
Three end devices are connected via two bridges.
The end devices 1 and 2 on the left side send frames via a total of three streams to end device 3 on the right side.
Additionally, we display the transmission times on the two network links shared by all streams.
On the first highlighted network link, the blue frame is transmitted immediately after the first green frame. 
On the second link there is a gap between the green and blue frame, even though we consider no-wait transmissions.
This is because the frame size of the blue stream is larger, leading to longer transmission delays.
This in turn, leads to a higher offset ($\phi$) for the yellow stream. 
If it was transmitted back to back on the first highlighted link, it would catch up at the second link and require queuing.

We denote the set of streams admitted within a single traffic plan as $\mathcal{S}$.
However, we expect changes in the system over time, which is reflected in streams leaving the system and new streams trying to join it. 
To this end, the traffic plan is iteratively updated and $\mathcal{S}_i$ represents the admitted streams at iteration $i$.
The streams trying to join the system in iteration $i + 1$ are collected in $\mathcal{S}^{\text{add}}_{i+1}$.
Meanwhile, $\mathcal{S}^{\text{del}}_{i+1}$ holds the admitted streams that are leaving the system in iteration $i +1$. 
Streams from $\mathcal{S}^{\text{add}}_{i+1}$ the scheduler cannot admit are rejected $\mathcal{S}^{\text{rej}}_{i+1} \subseteq \mathcal{S}^{\text{add}}_{i+1}$.
Hence, the streams admitted at iteration $i+1$ can be denoted as $\mathcal{S}_{i+1} = \mathcal{S}_i \setminus \mathcal{S}^{\text{del}}_{i+1} \cup \mathcal{S}^{\text{add}}_{i + 1} \setminus \mathcal{S}^{\text{rej}}_{i+1}$.
Note that rejecting previously admitted streams is prohibited.
Every admitted stream needs to be served until its removal is requested.
However, an admitted stream can be reconfigured, i.e., be assigned a different route or phase as long as its end-to-end delay guarantees are met.


\section{Fundamentals of Conflict-Graph-based Traffic Planning}
\label{sec:preliminaries}

The next section covers conflict-graph-based traffic planning.
We start by formalizing the traffic planning problem, followed by a definition for conflict graphs.
Further, we show how to model the traffic planning problem as a conflict graph and why the graph representation is beneficial for the scheduling problem.

\subsection{Traffic Planning for Time-triggered Streams}
\label{sec:preliminaries:traffic_planning_problem}

In this work, we address the problem of planning the routes and transmission times for a set of periodic, time-triggered streams $\mathcal{S}$ within a time-sensitive network $\mathcal{N}$ over the course of a (potentially unknown) number of iterations $\mathcal{I}$.
To this end, we need to ensure spatial or temporal isolation for all frames, i.e., no two frames are transmitted via the same egress port at the same time.
This is achieved by selecting suitable routes $\pi$ and phases $\phi$ for every stream so that their frames do not interfere with each other.

\begin{figure}
	\centering
	\includegraphics[width=0.45\linewidth]{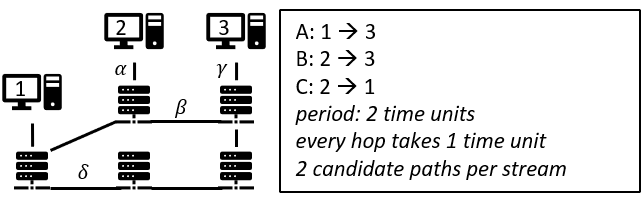}
	\caption{Example network with 3 simplified time-triggered streams. Conflicts happen at the network links $\alpha$, $\beta$, $\gamma$, and $\delta$.}
	\label{fig:networkExample}
\end{figure}

The optimization objective of the traffic planning is to minimize the number of rejected streams in every iteration, or formally: $\forall i \in \mathcal{I}: \min \left |\mathcal{S}^{\text{rej}}_i\right |$.
Figure~\ref{fig:networkExample} shows a small network with three end devices and three simplified streams ($A$, $B$, and $C$).
These streams can interfere at the network links $\alpha$, $\beta$, $\gamma$, and $\delta$.
To admit all of them, we need to select an appropriate route and phase for each stream, which we will do via a conflict graph-based scheduling approach.

\subsection{Conflict Graph}
\label{sec:preliminaries:conflict_graph}

\begin{figure}
	\centering
	\includegraphics[width=0.45\linewidth]{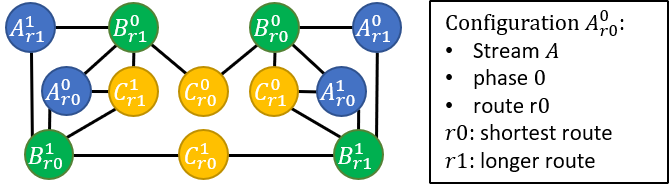}
	\caption{Conflict graph for the network and streams depicted in Figure~\ref{fig:networkExample}.}
	\label{fig:cgExample}
\end{figure}

The conflict graph is an undirected, vertex-colored graph $\mathcal{G}_\mathcal{C}=(\mathcal{V},\mathcal{E},\mathcal{C})$, where $\mathcal{V}$ is the vertex set, $\mathcal{E} \subseteq \mathcal{V}\times\mathcal{V}$ the set of undirected edges, and $\mathcal{C}$ the set of colors.
Further, let the function $c(v): \mathcal{V} \rightarrow \mathcal{C}$ map vertices to their respective color.

The vertices in $\mathcal{G}_\mathcal{C}$ model possible scheduling \emph{configurations} as $\pi-\phi$ (route and phase) combinations for a specific stream that do not violate the required end-to-end delay.
An example for the network problem from Figure~\ref{fig:networkExample} is shown in Figure~\ref{fig:cgExample}.
The vertex color denotes the configuration's stream, i.e., the vertex on the upper left: $A^1_{r1}$ , belongs to stream $A$ (blue color and the A label for clarity).
The indices in the vertex label denote the route and phase.
In this example, we use $r0$ to denote the shortest route between source and destination, and $r1$ to denote the other loop-less route.
For the phases, we consider 0 and 1 due to the period of 2.
For simplicity, we ignore the period = deadline assumption from the system model in this example and assume it takes exactly one time unit to forward a frame a single hop.
Two configurations are in conflict with each other when they are not spatially or temporally separated, making them mutually exclusive.
For example, configurations of $B$ and $C$ conflict if they have the same phase at link $\alpha$.
Edges between two configurations symbolize these conflicts.
Note that we need to check all frames of both streams withing a hypercycle and if any two frames conflict, these two configurations are in conflict.
All configurations in $\mathcal{G_C}$ need to be checked for conflicts pairwise.
It is possible to create subgraphs by removing vertices and their adjacent edges (or not creating them in the first place) to have a smaller conflict graph that ignores some possible scheduling options.
This becomes important for solving the graph problem efficiently.

\subsection{Conflict Graph-based Scheduling}
\label{sec:preliminaries:ics}
\label{sec:preliminaries:cg-based_scheduling}

To find a valid traffic plan for the network problem discussed in Section~\ref{sec:preliminaries:traffic_planning_problem}, we need to find appropriate routes and phases for every stream, so that they do not interfere with each other.
Thereby, the conflict graph representation helps us to abstract the domain specific network scheduling problem and can instead solve a generic graph problem: the independent colorful set problem.
The goal is to find a set of independent vertices, i.e., vertices that are not directly connected by an edge, in a vertex-colored graph $\mathcal{G_C}$.
In Section~\ref{sec:preliminaries:conflict_graph} we described how each vertex in the conflict graph represents one configuration for a specific network stream.
Links between vertices meant that these two configurations are mutually exclusive.
Each stream was mapped to a unique color, so that an independent colorful set can be translated into a valid schedule for the network problem.
The Independent Colorful Set problem can be solved using existing approaches, like the Greedy Flow Heap heuristic \cite{Falk2022}.

In a dynamic system, each change in the stream set $\mathcal{S}_i$ results in a new conflict graph. 
Especially adding new streams $\mathcal{S}^\text{add}_{i+1}$ requires adding now vertices (and colors) to $\mathcal{G_C}$.
This means, for every iteration $i$ the independent colorful set problem needs to be solved again.

\section{Problem Statement}
\label{sec:problem-statement}
\label{sec:problem-statement:problem-statement}

While there already exist algorithms to find an independent colorful set in $\mathcal{G_C}$ \cite{Falk2020,Falk2022}, the transformation from the traffic planning problem into the ICS problem, i.e., the creation of the conflict graph, is not yet efficient.
This issue is intensified in dynamic systems, when the conflict graph has to be updated frequently.

Deleting configurations of removed or rejected streams is straightforward. 
The major challenge is to append new configurations. 
Since this is our major concern, we refer to all activities involved in the construction and modification of the graph as \textbf{expansion}.
Adding a single configuration (vertex) requires a conflict check against every existing configuration in $\mathcal{G_C}$.
This leads to a linear increase in complexity as the conflict graph grows.
Hence, a \enquote{good} conflict graph is small while still yielding a colorful set containing all or at least most new streams.

In practice, however, admitting all streams can be impossible within seconds or minutes, especially if the conflict graph is growing large.
Therefore, we relax our understanding of a \enquote{good} conflict graph by specifying that, for any given number of configurations $\overline{\mathcal{V}}$, it should admit as many streams as possible, while keeping the quality-of-service guarantees of previous streams.
The number of configurations is defined as $\overline{\mathcal{V}} = \text{cps} \times \left | \mathcal{S}_i \setminus \mathcal{S}^{\text{del}}_{i+1} \cup \mathcal{S}^{\text{add}}_{i+1} \right |$, i.e., we consider an average budget of cps configurations per stream of the current iteration.
This includes the old streams from previous iterations $\mathcal{S}_{0..i}$ as well as the new streams from $\mathcal{S}_{i+1}$.
Good values for cps will be investigated in the evaluation.
Summarized, our goal is to create a conflict graph of limited size which yields a colorful independent set with as many colors (streams) as possible.

Note that, it is necessary to perform the conflict graph expansion repeatedly in dynamic systems. 
Similar to the independent colorful set, the expansion has to be performed for every iteration $i$, which makes it even more desirable to have small conflict graphs, so that the expansion times remain feasible. 
The streams admitted in earlier iterations need to be part of all later solutions until they are removed (part of $\mathcal{S}^\text{del}_{i+1}$).

\section{Conflict Graph Expansion}
\label{sec:conflict_graphs}

Now we will discuss how to expand (create or update) the conflict graph.
We introduce our classification scheme for the expansion strategies with regard to the $\pi-\phi$--space enumeration and configuration distribution.
Next, we describe the approach by Falk et al. \cite{Falk2022}, and discuss it in the context of the proposed classification. 
Finally, we provide a better route and phase enumeration, followed by multiple distribution strategies.

We classify conflict graph expansion strategies based on two orthogonal features: the route-phase-enumeration and the configuration distribution between the streams.
When creating configurations for a stream, combinations of routes $\pi$ and phases $\phi$ need to be selected.
This selection can be \emph{deterministic}, i.e., it is performed in a structured, predictable way.
An example would be to increment $\phi$ by 1 for every configuration, and then to reset $\phi$ and increment $\pi$ when the maximum value on $\phi$ is reached.
Instead, the selection can also be \emph{randomized}, e.g., by selecting $\pi$ and $\phi$ randomly from the set of combinations that have not yet been selected.
Note that different selection strategies can be applied to the route and the phase dimension, e.g., by picking the same number of configurations for every route $\pi$, but selecting their $\phi$ values randomly.

For the configuration distribution, we distinguish between \emph{homogeneous} and \emph{heterogeneous} approaches.
In a homogeneous approach, every stream receives the same number of configurations, while in a heterogeneous approach, some streams receive fewer configurations than others.
A heterogeneous distribution can be desirable if not every stream is equally \enquote{hard} to schedule.
With heterogeneous distributions, these \enquote{hard} streams can either receive more configurations to find a solution that accommodates them, or less, so that \enquote{easy} streams have more options to be placed around the \enquote{hard} streams.
We will cover heterogeneous distributions and their intentions in more detail in Section~\ref{sec:cg-expansion:heterogeneous}.

The conflict graph expansion described by Falk et al. uses a deterministic $\pi-\phi$--space enumeration \cite{Falk2022}.
It starts with $\pi=\phi=0$, and then increments the route $\pi$ by 1. 
As soon as every candidate route has a configuration with $\phi=0$, $\phi$ is increased by a fixed amount $\Delta$ and the procedure is repeated.
The value of $\Delta$ is defined as the 75\%-fractile of the transmission times.
Consequently, a shift by $\Delta$ solves many conflicts.
In the approach described by Falk et al., configurations are added to both new and old streams, with the objective to equalize the numbers over time.
This makes their approach effectively a homogeneous distribution strategy.
We will not expand streams from previous iterations, since this leads to extremely large conflict graphs.

\subsection{Route and Phase Enumeration}
\label{sec:conflict_graphs:random_phase}

A common problem with deterministic $\pi-\phi$--space enumeration is that it leads to recurring conflicts.
For example, if two routes A and B are in conflict for a given phase $\phi$, then they will have the same conflict for any other value of $\phi$.
This problem is intensified if multiple routes are in conflict, i.e., they form a clique in the conflict graph.
Changing the route $\pi$ is one way to mitigate the issue.
However, similar problems may arise on other candidate routes.

Therefore, we propose to randomize $\phi$.
This ensures that, when adding more configurations, they are less likely to replicate previous conflicts.
The routes, on the other hand, are not randomized, which means that every candidate route (within a stream) will receive a similar number of configurations.
The rationale behind this is that the routing already pursues certain objectives, such as having short but diverse routes \cite{Geppert2023}.
Note that these objectives cannot be met by all candidate routes if the candidate route pool becomes too large.
Therefore, a small number of candidate routes is beneficial.

Based on these considerations, we define the following conflict graph expansion workflow:
For each new stream, its candidate routes are computed, and the $\pi-\phi$ combinations to be added to the conflict graph are determined.
Thereby, we randomly select phases for every candidate route.
The same number of phases is given to every candidate route of a stream.
The total number of configurations that are created for a stream is determined either by a homogeneous or by a heterogeneous distribution strategy used alongside this $\pi-\phi$--space enumeration.

\subsection{Heterogeneous Randomized Phase Conflict Graph Expansion}
\label{sec:cg-expansion:heterogeneous}

The heterogeneous distribution strategies are responsible for determining the number of configurations the $\pi-\phi$--space enumeration can create for each stream.
We call these numbers the stream budgets.
These per stream budgets are derived from the total number of configurations $\overline{\mathcal{V}}$ and a base budget $\alpha$.
The base budget ensures that each stream has at least $\alpha$ configurations.
In our test setting, repeated measurements showed that a value of $\alpha=5$ yields good results.
As system updates are performed iteratively, we also need to keep previous expansions in mind.
Therefore, at iteration $i$, a heterogeneous distribution strategy operates on the remaining freely disposable configurations $R_i = \overline{\mathcal{V}} - \left |\mathcal{V} \backslash \{v \in \mathcal{V}: \text{col}(v) \in S^\text{del}_i\}\right | - \left |\mathcal{S}^{\text{add}}_i \right | \times \alpha$. 
We call them \enquote{freely disposable}, because all streams have already received at least as many configurations as specified in the base budget.
Thus, the remaining configurations can be distributed freely.

In our investigation of different heterogeneous distribution strategies, we experienced that it was more beneficial to create fewer configurations for \enquote{hard} streams, so that \enquote{easier} streams had the flexibility to be placed around the \enquote{hard} streams.
This can be primarily ascribed to our use of the Greedy Flow Heap (GFH) heuristic \cite{Falk2022} to solve the ICS problem, which prioritizes streams with few configurations.
We created different heterogeneous distribution strategies, each with its own, but often related, interpretation of \enquote{hard} streams.
In the following, we present a strategy applying the networking metric \emph{traffic volume}, and two strategies using graph metrics of the conflict graph, making them domain independent: \emph{average degree}, and \emph{page rank}.

\subsubsection{Traffic Volume}
\label{sec:cg-expansion:heterogeneous:traffic}

Intuitively, a configuration will have more conflicts if the related stream sends more frames (low period) or if the frames require more time to be sent (large frame size).
Hence, a stream is considered to be \enquote{harder} when it has a higher data rate, denoted as traffic volume $\text{vol}(s) = \frac{\text{size}(s)}{\text{period}(s)}$ of a stream $s$.

Following the \emph{traffic volume} strategy, the budget $b_{\text{vol}}(s)$ of a stream assigns the streams a number of configurations that is inversely proportional to their traffic volume (cf. Equation~\ref{eq:exp:vol}).

\begin{equation}
	\label{eq:exp:vol}
	b_{\text{vol}}(s) = \frac{\overline{v} - \text{vol}(s)}{\overline{v} \times \left |\mathcal{S}^{\text{add}}_{i + 1}\right | - \sum_{s'\in \mathcal{S}^{\text{add}}_{i + 1}}\text{vol}(s')} \times R_i + \alpha
\end{equation}

The numerator denotes the difference in the traffic volume between $s$ and the stream with the largest possible traffic volume $\overline{v} = \frac{1500}{\min \text{period}(\mathcal{S})}$, which is the maximum transmission unit (simplified to \SI{1500}{\byte}) divided by the smallest period in all streams $\mathcal{S}$.
The denominator normalizes this difference with the aggregated differences of all streams in $\mathcal{S}^\text{add}_{i+1}$.
This returns the proportion of freely disposable configurations $R_i$ that will be assigned to the budget of $s$.
This proportion is multiplied with $R_i$ and incremented by the baseline number of configurations~$\alpha$.

\subsubsection{Average Degree}
\label{sec:cg-expansion:heterogeneous:degree}

The idea of the average degree strategy is that streams that have many conflicts (adjacent edges in $\mathcal{G}_\mathcal{C}$) are likely to be \enquote{hard} streams.
The average degree and page rank strategies require a small change in the overall conflict graph expansion.
Instead of computing the budgets first and then adding configurations to the conflict graph, we expand in two steps.
First, for every stream $\alpha$ configurations are added to $\mathcal{G_C}$, so that graph metrics are available for the new streams.
Then, the per stream budget is computed, without adding $\alpha$ to the budget again.

Let $\text{deg}(s)$ denote the average degree of all configurations (vertices) of $s$ in $\mathcal{G}_C$ after the first expansion step and let $\overline{deg}$ be the maximum average degree of the streams in $\mathcal{S}^{\text{add}}$.
Equation~\ref{eq:exp:avgdegree} specifies the calculation of the stream budget $b_\text{degree}$ for a given stream $s$.
We again determine the difference in the average degree between stream $s$ and the stream with the maximum average degree, and normalize it, before multiplying it with the number of remaining freely disposable configurations.

\begin{equation}
	\label{eq:exp:avgdegree}
	b_{\text{degree}}(s) = \frac{\overline{\text{deg}} - \text{deg}(s)}{\overline{\text{deg}} \times \left |\mathcal{S}^{\text{add}}_{i + 1}\right | - \sum_{s'\in \mathcal{S}^{\text{add}}_{i + 1}}\text{deg}(s')} \times R_i
\end{equation}

\subsubsection{Page Rank}
\label{sec:cg-expansion:heterogeneous:pagerank}

The average degree strategy only considers a single hop in $\mathcal{G}_C$, but not the \enquote{hardness} of the one-hop neighbors.
The page rank strategy instead considers multiple hops by using the page rank algorithm as \enquote{hardness} assessment (cf. Equation~\ref{eq:exp:pagerank}).
The page rank strategy requires the same two-step expansion as the average degree strategy.

\begin{equation}
	\label{eq:exp:pagerank}
	b_{\text{pr}}(s) = \frac{\overline{\text{sr}} - \text{sr}(s)}{\overline{\text{sr}} \times \left |\mathcal{S}^{\text{add}}_{i + 1}\right | - \sum_{s'\in \mathcal{S}^{\text{add}}_{i + 1}}\text{sr}(s')} \times R_i 
\end{equation}

We define a stream rank $sr(s) = \sum_{v \in \mathcal{V}: c(v)=s} pr(v)$, as the sum of the page ranks $pr(v)$ of all configurations $v$ of a stream $s$.
The highest stream rank is denoted by $\overline{sr}$.
The page rank approach is much more computationally expensive than any of the other strategies.
Therefore, we approximate the page rank by running 4 page rank iterations.

\subsubsection{Other Metrics}

We did implement and evaluate heuristics based on other metrics like the route length, period, or a hybrid strategy considering period and traffic volume.
Also, we investigated these strategies with inverted ordering, i.e., expanding \enquote{hard} streams.
In our evaluations those strategies did not have any relevant benefits and were often worse than the baseline.
Therefore, we will present only the heuristics described in Sections~\ref{sec:cg-expansion:heterogeneous:traffic}--\ref{sec:cg-expansion:heterogeneous:pagerank} (traffic volume, average degree, and page rank), as they consistently demonstrated superior performance compared to the heterogeneous baseline strategy.


\section{Evaluation}
\label{sec:eval}

In the following, we describe our evaluation setup, followed by the metrics used to evaluate our strategies.
Afterwards, we showcase the performance improvements of our new conflict graph creation strategies compared to the original one, as well as established scheduling heuristics, and evaluate the conflict graph based scheduling in the context of an energy grid.

\subsection{Evaluation Setup}
\label{sec:eval:environment}

All experiments were conducted on an Ubuntu 20.04 machine, equipped with two AMD EPYC 7401 24-Core processors and \SI{256}{\giga\byte} RAM.
For a more realistic execution environment, we partitioned the system resources and only used 8 CPUs (threads) per scheduling job. 
The average RAM requirement was within a few \SI{}{\giga\byte}.
The expansion and scheduling algorithms were implemented in C++ and compiled with GCC. 
The Greedy Flow Heap heuristic was used for scheduling \cite{Falk2022}, the modified version of Dijkstra's algorithm presented in \cite{Geppert2023} for routing, and we employed defensive and offensive planning.

The evaluated topologies were formed by bridges, with every bridge being connected to one end device.
We evaluated the following network topologies \cite{Newman2018,Waxman1988}:
\begin{itemize}
	\item \emph{Erd\"{o}s--R\'{e}yni} random (in the following called \enquote{random}) -- randomized network; each vertex pair has an edge based on a probability $p$. 
	Those networks have small diameters with multiple alternative routes of comparable length.
	\item \emph{Waxman} -- vertices are distributed randomly in a rectangular space. 
	Edges between two vertices are drawn with probabilities based on their Euclidean distance, favoring short distances.
	Such topologies that favor links to geographically close neighbors are more realistic, since typically, cables follow short paths between end-points.
	\item \emph{Ring} -- vertices are arranged in a circle, every vertex is connected to both immediate neighbors.
	Ring topologies are often used in safety-critical networks such as industrial networks to provide redundancy with a minimum number of additional links. 
	\item \emph{Grid} -- vertices are arranged in a square mesh. Every vertex is connected to its 2-4 closest neighbors (2 in the corners, 3 at the rim, 4 inside the grid).
	Compared to random graphs, such networks have a larger diameter since shortcuts to distant parts of the network are missing.
\end{itemize}

For a given network size of $n$ bridges, we created 10 network instances for the random and Waxman topologies to avoid bias towards a specific instance.
In contrast, there exists only a single instance of size $n$ for our ring and grid networks.
Additionally, we used multiple communication scenarios to prevent bias towards a specific stream set.
The number of scenarios was further increased for ring and grid topologies to offset the single network instance.
Considering network speed, we assumed \SI{1}{\giga\bit\per\second} network links, with a propagation delay of \SI{1}{\micro\second}, and bridges with a processing delay of \SI{4}{\micro\second}~\cite{Duerr2014}.

The stream sets to be scheduled were created by selecting source and destination devices at random.
Period and frame size were randomly chosen from a fix set of values.
Frame sizes range from 125--\SI{1500}{\byte}, thus allowing for small messages as well as almost MTU sized ones.
The periods range from hundreds of \SI{}{\micro\second} to a few \SI{}{\milli\second} \cite{IECSC65C/WG18}. 
The exact period and frame size sets depend on the experiment.
In each experiment, we evaluated at least five different traffic scenarios for topologies with multiple network instances, i.e., random and Waxman, and 20 for topologies with a single network instance, i.e., ring and grid.

Considering the candidate route set, we use two candidate routes per stream.
Especially for smaller conflict graphs, i.e., with lower cps values, two paths yielded better results than having no routing, while more than two paths could lead to worse schedules.
This is because short, divergent candidate routes are desirable, so that the network resources needed are low, due to the limited amount of hops, while bottlenecks in one route are not present in other routes, since they use different network links.
Hence, low cps values yield a good saturation of offsets for two candidate routes, while additional candidate paths need more configurations to be effective.

\subsection{Benchmark Algorithms}
\label{sec:eval:banchmark-algorithms}
Within our evaluation, we use several benchmark algorithms. All of them allow for offensive and defensive planning.

Additionally, we implemented the very fast \textbf{Greedy} heuristic from \cite{Raagaard2017a}.
It considers only the shortest path, but can use all 8 queues available in a TSN bridge.
Hence, it is not a no-wait approach.

Finally, we added the \textbf{Heuristic List Scheduler (HLS)} \cite{Pahlevan2019} from Pahlevan et al. to our evaluation.
HLS is a no-wait approach that considers all possible paths a stream can take.
However, this is not scalable since there the number of possible paths quickly leads to crashes due to insufficient memory ($>$\SI{256}{\giga\byte}) or very long runtimes.
Therefore, we only consider the 100 shortest paths using yen's k-shortest path algorithm on larger topologies \cite{Yen1971}. 
Also, originally HLS was designed for use cases with more complex applications, i.e, with a dedicated application graph dictating precedence constraints in the communication streams.
Since we do not need the application graph capability, it was left out.

We do not present a comparison against any solver based approach, e.g., Constraint Programming.
This is because in preliminary evaluations these approaches did not solve our problem sizes within any reasonable time.

\subsection{Performance Metrics}

In this section, we discuss the metrics used to evaluate the performance of the different strategies.
The \emph{number of rejected streams} indicates how well the optimization goal of admitting as many streams as possible is met.
The lower the number of rejected streams, the better.

The \emph{total time} denotes the whole runtime for a scheduling iteration and consists of two parts:
First, the \emph{expansion time} measures how much time was spent expanding the conflict graph (only applicable on the conflict graph based approaches).
Second, the \emph{scheduling time}, denoting the time needed to solve the ICS problem on a given conflict graph or alternative type of computation for the other benchmarked algorithms.

The \emph{number of conflicts} denotes the number of edges in the conflict graph, i.e. the size of $\mathcal{G}_\mathcal{C}$.
It is generally easier to find a complete set admitting all streams in a graph with fewer edges, as it becomes easier to select low degree configurations with low impact on the remaining solution space.
Thus, the number of edges is an indication for the computational effort it takes to find a complete set.
However, it is not guaranteed that a complete set will be found or even exists.
In addition to the lower computational effort, smaller graphs are desirable in terms of their lower memory requirement, since the memory needed scales with the number of edges.

\subsection{Results}

In the following, we first show the improvements in the conflict graph creation compared to the previous methods.
This includes the switch from deterministic phase enumeration to randomized phase enumeration, and the heterogeneous distribution strategies compared to the homogeneous strategy.
Thereby, we investigate dynamic scenarios, the impact of harmonic and non-harmonic periods, scalability, and benchmark them against other heuristics.
Finally, we showcase the suitability of the conflict-graph-based scheduling with our scalable expansions in an energy metering scenario.

\subsubsection{Randomized Phase Enumeration}
\label{sec:eval:randomized_phase}

\begin{figure}
	\centering
	\subfloat[Scheduling success\label{fig:phase-expansion-waxman-49:rejected}]{
		\centering
		\includegraphics[trim={4mm, 6mm, 4mm, 4mm}, clip, width=0.3\linewidth]{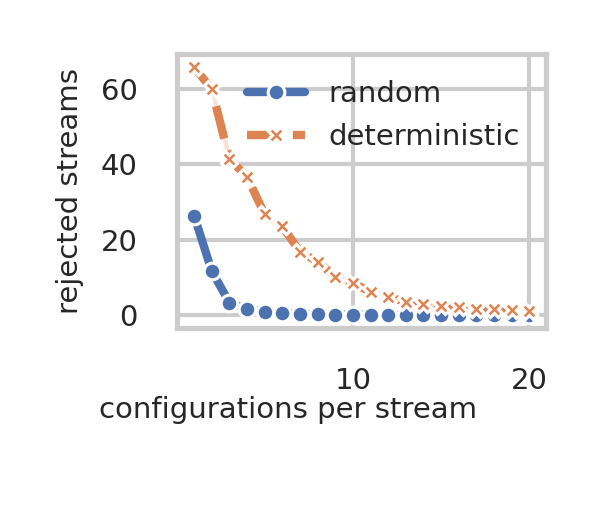}
	}
	\subfloat[Conflict Graph size\label{fig:phase-expansion-waxman-49:cg-size}]{
		\centering
		\includegraphics[trim={4mm, 6mm, 4mm, 4mm}, clip, width=0.3\linewidth]{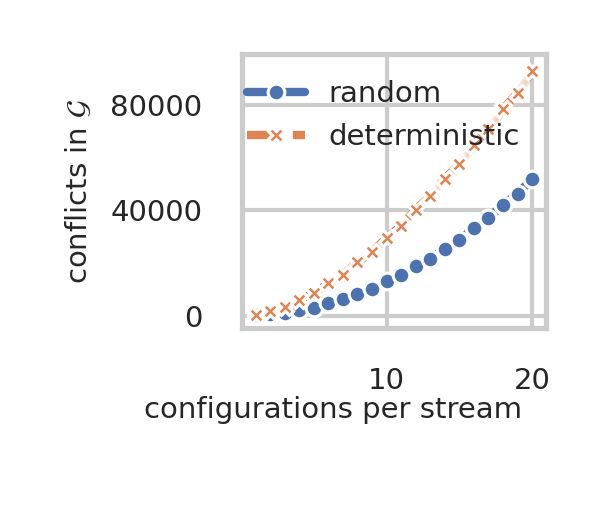}
	}
	\caption{Comparison of randomized vs deterministic phase enumeration in a Waxman network with 49 bridges.}
	\label{fig:phase-expansion-waxman-49}
\end{figure}

In the following, we compare the randomized enumeration we proposed in Section~\ref{sec:conflict_graphs:random_phase} to the deterministic enumeration from \cite{Falk2022}.

We evaluate the scheduling on different Waxman networks with 49 bridges each and add 100 streams with non-harmonic periods.
Other topologies yield similar results.
We varied the number of configurations per stream (cps), ranging from 1 to 20.
We only ran a single iteration without dynamic updates, i.e. we performed offline scheduling, to highlight the impact of $\text{cps}$ on randomized and deterministic phase enumeration. 

Figure~\ref{fig:phase-expansion-waxman-49:rejected} shows the number of rejected streams.
It can be seen that the deterministic approach rejects many more streams initially.
As cps and thus $\overline{\mathcal{V}}$ grow, both approaches reject fewer streams.
However, the randomized approach admits all streams way sooner.

We also investigated the number of conflicts in the conflict graph (number of edges).
The deterministic approach starts with four times the number of edges, and increases much more steeply with the number of cps (cf. Figure~\ref{fig:phase-expansion-waxman-49:cg-size}).
The higher number of conflicts makes it less likely to admit all streams and increases the scheduling time.
There is no major difference in the expansion time (not shown).

In summary, we see that the randomized phase enumeration performed much better in terms of rejected streams and produces a conflict graph that is only half the size of the deterministic graph.
When using harmonic periods, the conflict graph shrank to about one third.
The following evaluations will only use the randomized enumeration unless stated otherwise.

\subsubsection{Scheduling Dynamic Systems}
\label{sec:eval:dynamic}

Next, we take a deeper look at the distribution strategies in dynamic systems.
We investigated multiple network topologies, by setting up a certain number of initial streams in an initialization step, followed by 20 iterations representing the dynamic system updates.
In each of the 20 iterations 20 streams were removed, and 20 new streams joined the system.
The frame sizes and periods were randomly selected from \{125, 250, 500, 750, 1000, 1500\}~\SI{}{\byte} and \{300, 400, 500, 1200, 1500, 2000\}~\SI{}{\micro\second}.
The configuration limit $\overline{\mathcal{V}}$ is again defined in terms of cps, but this time using a fixed value of $\text{cps}=50$.

We evaluated our heterogeneous distribution methods against the homogeneous distribution (all of them utilizing the GFH algorithm \cite{Falk2022}), the Greedy scheduling, and HLS.
Note that the original HLS does not run on this topology because there are too many possible routes. 
Hence, we had to bound the routing (cf. Section~\ref{sec:eval:banchmark-algorithms}).

\begin{table}
	\renewcommand{\arraystretch}{1.08}
	\begin{tabularx}{\linewidth}{l|p{1.5cm}p{1.5cm}p{1.5cm}p{1.5cm}}
		Strategy & rejected streams & expansion time [s] & solving time [s] & total time [s] \\
		\multicolumn{5}{c}{Waxman 200 streams ($\sim$\SI{1860}{\mega\bit\per\second})}\\
		\cline{1-5}HLS & 0.34$\pm$1.22 & - & 0.91$\pm$1.19 & 0.91$\pm$1.19\\
		Greedy & 3.29$\pm$3.73 & - & 0.00$\pm$0.00 & 0.01$\pm$0.00\\
		\cline{1-1}
		homogeneous & \textbf{0.31$\pm$1.13} & 0.15$\pm$0.05 & 0.03$\pm$0.02 & 0.20$\pm$0.07\\
		degree & 0.36$\pm$1.12 & 0.11$\pm$0.03 & 0.03$\pm$0.01 & 0.16$\pm$0.04\\
		page rank & 0.35$\pm$1.08 & 0.17$\pm$0.05 & 0.03$\pm$0.01 & 0.22$\pm$0.06\\
		traffic volume & \textbf{0.31$\pm$1.13} & 0.15$\pm$0.05 & 0.03$\pm$0.02 & 0.20$\pm$0.07\\
		\multicolumn{5}{c}{Waxman 400 streams ($\sim$\SI{3300}{\mega\bit\per\second})}\\
		\cline{1-5}HLS & \textbf{1.62$\pm$2.36} & - & 3.77$\pm$3.02 & 3.77$\pm$3.02\\
		Greedy & 9.63$\pm$4.80 & - & 0.01$\pm$0.00 & 0.01$\pm$0.00\\
		\cline{1-1}
		homogeneous & 2.10$\pm$3.10 & 0.24$\pm$0.06 & 0.10$\pm$0.05 & 0.39$\pm$0.09\\
		degree & 2.12$\pm$2.96 & 0.17$\pm$0.04 & 0.08$\pm$0.03 & 0.28$\pm$0.06\\
		page rank & 2.11$\pm$2.90 & 0.40$\pm$0.08 & 0.08$\pm$0.03 & 0.51$\pm$0.10\\
		traffic volume & 2.07$\pm$3.07 & 0.23$\pm$0.06 & 0.10$\pm$0.05 & 0.38$\pm$0.09\\
		\multicolumn{5}{c}{grid 200 streams ($\sim$\SI{1860}{\mega\bit\per\second})}\\
		\cline{1-5}HLS & \textbf{0.00$\pm$0.00} & - & 0.47$\pm$0.69 & 0.47$\pm$0.69\\
		Greedy & 0.36$\pm$0.90 & - & 0.00$\pm$0.00 & 0.00$\pm$0.00\\
		\cline{1-1}
		homogeneous & 0.00$\pm$0.02 & 0.10$\pm$0.02 & 0.02$\pm$0.02 & 0.14$\pm$0.03\\
		degree & 0.00$\pm$0.03 & 0.08$\pm$0.01 & 0.02$\pm$0.01 & 0.12$\pm$0.02\\
		page rank & 0.00$\pm$0.04 & 0.13$\pm$0.02 & 0.02$\pm$0.01 & 0.17$\pm$0.03\\
		traffic volume & \textbf{0.00$\pm$0.00} & 0.10$\pm$0.02 & 0.02$\pm$0.02 & 0.14$\pm$0.03\\
		\multicolumn{5}{c}{grid 400 streams ($\sim$\SI{3660}{\mega\bit\per\second})}\\
		\cline{1-5}HLS & \textbf{0.06$\pm$0.25} & - & 1.65$\pm$2.37 & 1.66$\pm$2.37\\
		Greedy & 5.27$\pm$3.49 & - & 0.01$\pm$0.00 & 0.01$\pm$0.00\\
		\cline{1-1}
		homogeneous & 0.10$\pm$0.32 & 0.18$\pm$0.03 & 0.07$\pm$0.04 & 0.28$\pm$0.06\\
		degree & 0.15$\pm$0.40 & 0.14$\pm$0.02 & 0.05$\pm$0.03 & 0.22$\pm$0.04\\
		page rank & 0.17$\pm$0.41 & 0.37$\pm$0.04 & 0.05$\pm$0.03 & 0.45$\pm$0.06\\
		traffic volume & 0.09$\pm$0.31 & 0.17$\pm$0.03 & 0.07$\pm$0.04 & 0.27$\pm$0.05\\
		\multicolumn{5}{c}{random 200 streams ($\sim$\SI{1860}{\mega\bit\per\second})}\\
		\cline{1-5}HLS & \textbf{0.00$\pm$0.00} & - & 0.32$\pm$0.22 & 0.32$\pm$0.22\\
		Greedy & 0.06$\pm$0.34 & - & 0.00$\pm$0.00 & 0.00$\pm$0.00\\
		\cline{1-1}
		homogeneous & \textbf{0.00$\pm$0.00} & 0.07$\pm$0.01 & 0.00$\pm$0.01 & 0.09$\pm$0.02\\
		degree & \textbf{0.00$\pm$0.00} & 0.06$\pm$0.01 & 0.01$\pm$0.01 & 0.09$\pm$0.02\\
		page rank & \textbf{0.00$\pm$0.00} & 0.09$\pm$0.01 & 0.01$\pm$0.01 & 0.11$\pm$0.02\\
		traffic volume & \textbf{0.00$\pm$0.00} & 0.07$\pm$0.01 & 0.00$\pm$0.01 & 0.09$\pm$0.02\\
		\multicolumn{5}{c}{random 600 streams ($\sim$\SI{5380}{\mega\bit\per\second})}\\
		\cline{1-5}HLS & \textbf{0.01$\pm$0.09} & - & 1.03$\pm$2.44 & 1.03$\pm$2.44\\
		Greedy & 4.28$\pm$2.90 & - & 0.01$\pm$0.00 & 0.01$\pm$0.00\\
		\cline{1-1}
		homogeneous & 0.06$\pm$0.29 & 0.19$\pm$0.04 & 0.09$\pm$0.04 & 0.32$\pm$0.07\\
		degree & 0.09$\pm$0.33 & 0.15$\pm$0.03 & 0.07$\pm$0.04 & 0.25$\pm$0.05\\
		page rank & 0.09$\pm$0.33 & 0.42$\pm$0.07 & 0.07$\pm$0.04 & 0.52$\pm$0.09\\
		traffic volume & 0.06$\pm$0.29 & 0.18$\pm$0.04 & 0.09$\pm$0.04 & 0.31$\pm$0.07\\
		\multicolumn{5}{c}{ring 200 streams ($\sim$\SI{1860}{\mega\bit\per\second})}\\
		\cline{1-5}HLS & 7.14$\pm$2.79 & - & 0.08$\pm$0.01 & 0.08$\pm$0.01\\
		Greedy & 6.91$\pm$5.05 & - & 0.01$\pm$0.00 & 0.01$\pm$0.00\\
		\cline{1-1}
		homogeneous & \textbf{6.83$\pm$3.36} & 0.43$\pm$0.07 & 0.05$\pm$0.01 & 0.54$\pm$0.09\\
		degree & \textbf{6.83$\pm$3.36} & 0.29$\pm$0.06 & 0.04$\pm$0.01 & 0.37$\pm$0.07\\
		page rank & 6.85$\pm$3.34 & 0.41$\pm$0.07 & 0.04$\pm$0.01 & 0.49$\pm$0.08\\
		traffic volume & 6.84$\pm$3.43 & 0.41$\pm$0.07 & 0.05$\pm$0.01 & 0.53$\pm$0.09\\
		\cline{1-5}
		Strategy & rejected streams & expansion time [s] & solving time [s] & total time [s] \\
	\end{tabularx}
	\caption{Dynamic scheduling on different networks and varying network utilization. Each
		 network update consisted of removing 20 streams and scheduling 20 new streams.}
	\label{tab:dynamic-results}
\end{table}

Table~\ref{tab:dynamic-results} displays the results of the dynamic scheduling iterations.
In this evaluation Greedy is almost constantly outperformed by all other strategies in terms of rejected streams, but runs much faster than any other strategy.
On the ring topology the Greedy approach admits more traffic, i.e., more data is transmitted, than for the other strategies while rejecting a similar number of streams.
This seems to be a side effect of Greedy scheduling the high traffic streams first.
In some cases HLS rejects the fewest streams, at the cost of the longest runtime.
However, the conflict-graph-based strategies are always close in terms of rejected streams if not even better.
When the number of streams increases, the runtime of HLS exceeds one second, while all other strategies are still far below.
Again, the only exception is the ring topology which does not have relevant routing options, thus leading to short runtimes in HLS.

When comparing the conflict-graph-based strategies, they reject streams comparable to HLS while the runtime is significantly better on all topologies except for ring.
Further, all conflict-graph-based strategies perform quite similar in this evaluation.
There are small, but insignificant, differences in the number of rejected streams and more diversity in the expansion time.
Page rank has an especially high expansion time due to the computational overhead.
However, page rank and degree also have much fewer conflicts (edges) in the conflict graph, leading to shorter scheduling times.

To summarize, the conflict-graph-based strategies perform well, and operate fast, even when the number of streams increases.
So they can react to system changes in a variety of topologies and even if we have non-harmonic periods.
However, they become more effective with harmonic periods, as we will show in the following.

\subsubsection{Harmonic Periods}

Harmonic stream periods are widely used in industrial applications \cite{Mohaqeqi2018}.
Therefore, we also investigate our scheduling approach with harmonic periods.
In this evaluation we performed 10 iterations on a 7x7 grid network.
We add 700 streams in the first (initial) iteration, while all following iterations, simulating dynamic system updates, add 50 streams and remove 25 streams, overcrowding a saturated network.
The stream's frame sizes and periods were randomly chosen from $\{125, 250, 500, 750, 1000, 1500\}$~\SI{}{\byte} and $\{250, 500, 1000, 2000\}$~\SI{}{\micro\second}.

\begin{figure}
	\centering
	\includegraphics[trim={5mm, 5mm, 4mm, 4.5mm}, clip, width=0.4\linewidth]{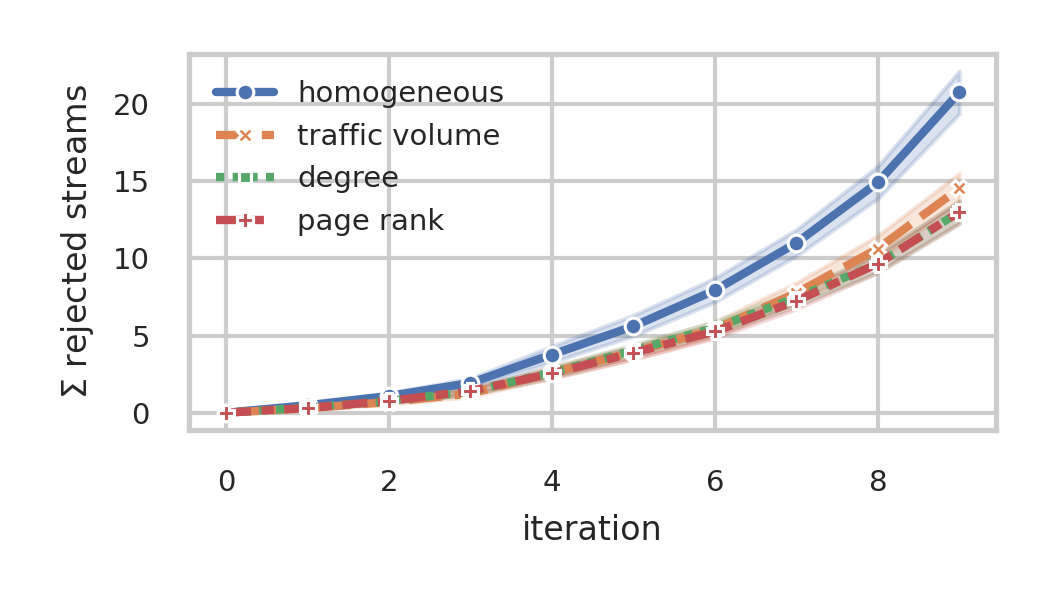}
	\caption{Stream rejection on a 7x7 grid when the streams have harmonic periods.}
	\label{fig:rejected-grid-7x7-scenario49grid700}
\end{figure}

The results are shown in Figure~\ref{fig:rejected-grid-7x7-scenario49grid700}. 
Note that, we did not include Greedy and HLS in the plot, because both of them performed much worse than the conflict-graph-based approaches which would make it harder to see the differences within our strategies.
After 10 iterations, Greedy rejected more than 350 streams, HLS more than 70, while all conflict-graph-based approaches stay below 25.
Meanwhile, the runtime of HLS was substantially longer than any other (>\SI{10}{\second} vs <\SI{1.5}{\second} for the dynamic updates).
We experienced the same behavior for other topologies.

The heterogeneous distribution strategies perform better than the homogeneous expansion.
Additionally, all heterogeneous strategies have smaller conflict graphs (fewer edges), leading to shorter scheduling times.
Degree and traffic volume also have slightly shorter expansion times, and therefore, outperform the homogeneous benchmark in every metric.
On random topologies we had similar results, while the differences in rejected streams are negligible on Waxman and ring topologies of this size.

\subsubsection{Scheduling in Large-scale Networks} 
\label{sec:eval:scaling}

Next, we investigate the performance of our heterogeneous distribution strategies in larger networks and focus especially on the impact the conflict graph size has by varying the cps value.
The cps value ranges from 30 to 80 with a step size of 10.
We conduct the evaluation on Waxman topologies with 256 bridges, frame sizes of $\{125, 250, 500, 750, 1000, 1500\}$~\SI{}{\byte} and periods of $\{250, 500, 1000, 2000\}$~\SI{}{\micro\second}.
In a single scheduling step we try to add 9000 streams into an empty system.
We ran only one large iteration instead of dynamic updates, since this way we have computationally complex and time-consuming computation which is more susceptible to scaling effects.
Greedy and HLS are not shown in the plot, since Greedy rejected more than 600 streams and HLS does not scale sufficiently. 

\begin{figure*}
	\subfloat[Scheduling success\label{fig:flexible-cps:rejected-waxman-256-scenarioflexiblecps}]{
		\centering
		\includegraphics[trim={4mm, 4mm, 4mm, 4mm}, clip, width=0.26\linewidth]{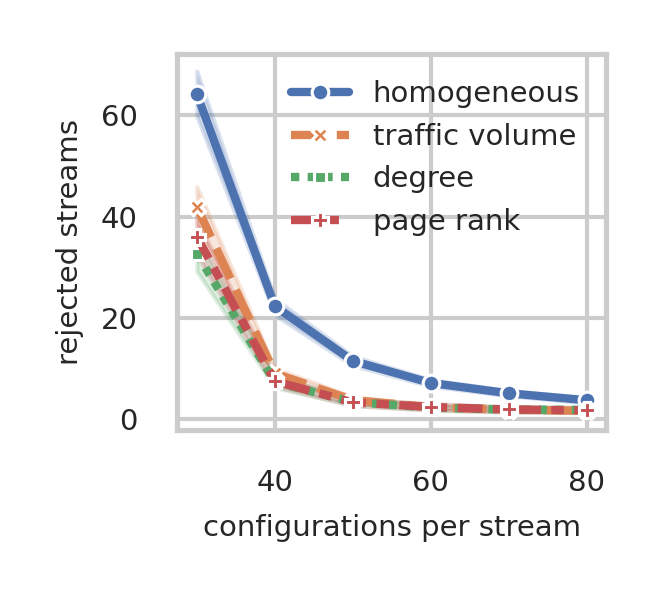}
	}
	\hfill
	\subfloat[Expansion time\label{fig:flexible-cps:expansion-time-waxman-256-scenarioflexiblecps}]{%
		\centering
		\includegraphics[trim={4mm, 4mm, 4mm, 4mm}, clip, width=0.26\linewidth]{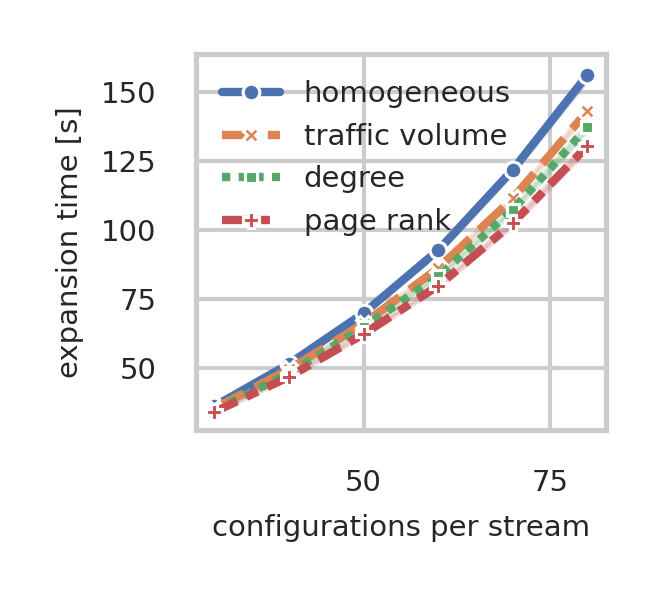}
	}
	\hfill
	\subfloat[Scheduling time\label{fig:flexible-cps:solving-time-waxman-256-scenarioflexiblecps}]{%
		\centering
		\includegraphics[trim={4mm, 4mm, 4mm, 4mm}, clip, width=0.26\linewidth]{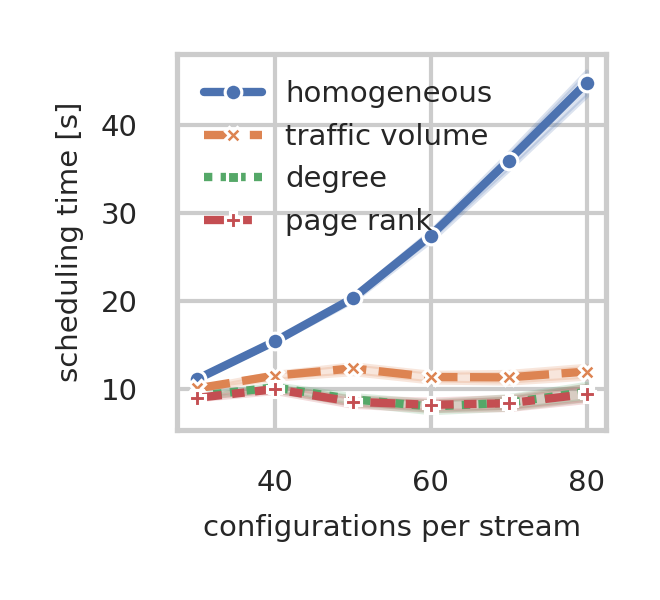}
	}
	\caption{Scheduling 9000 streams in a Waxman network with 256 bridges and variable cps}
	\label{fig:flexible-cps}
\end{figure*}

Figure~\ref{fig:flexible-cps} shows the number of rejected streams, conflict graph size, and scheduling time.
Our heterogeneous distribution strategies clearly outperform the homogeneous distribution strategy, not only by rejecting substantially fewer streams, thus admitting all streams with lower cps, but also with shorter runtimes.
Thereby, the difference in the scheduling time is especially apparent in Figure~\ref{fig:flexible-cps:solving-time-waxman-256-scenarioflexiblecps}, where our approaches remain almost identical in terms of runtime, while homogeneous' solving time increases substantially.
When also considering the expansion time (Figure~\ref{fig:flexible-cps:expansion-time-waxman-256-scenarioflexiblecps}), we first see that the all approaches have similar expansion times, but homogeneous is slightly slower.
Second, while the expansion time is still the dominant factor, the difference between solving and expansion time is less than an order of magnitude.
For cps values $<30$ all strategies rejected a lot of streams, so that these low cps values are not viable for this problem size.

\subsubsection{Use Case: Advanced Metering Infrastructure}

Next, we test our conflict graph expansion strategies in an advanced metering infrastructure (AMI) of a synthetic, yet realistic smart grid.
AMI enables two-way communication between energy consumers, producers, and providers~\cite{Gungor2013}.

\paragraph{Network Topology}

We use the IEEE 300--Bus System\footnote{\url{https://electricgrids.engr.tamu.edu/electric-grid-test-cases/ieee-300-bus-system/}} grid, a synthetic power grid case developed by the IEEE Test Systems Task Force.
In order to implement our scheduling approach we need to transform the power grid network into a communication network.
First, we augment the power grid topology by adding bridges at every transformer and bus.
Then we place end devices at loads, generators, and transformers. 
The network links follow the topology of the grid \cite{Zhu2019}.
This results in a communication network with 405 bridges, 371 end devices, and 886 network links.

\paragraph{Communication Scenario}

Similar to the energy grid topology, traffic patterns of AMI applications are critical information and therefore not publicly available.
Hence, we define a scenario that seeks to replicate the communication of an AMI application.
We use the 300-Bus System to model a high voltage backbone grid with local and backhaul concentrators.
The local concentrators collect smart meter information from low and medium voltage grids, e.g., from a residential area \cite{Bian2014}, which is not part of the topology.
For the backhaul concentrators, we randomly select three load nodes from the grid, since the IEEE 300-Bus System has three geographical areas.
Each remaining end device acts as a local concentrator and sends data streams to a nearby backhaul concentrator.
Since two-way communication is a central requirement to AMI, we return a stream to the local concentrator for each of its outgoing streams.
Considering the increasing prevalence of smart meters \cite{Kerai2022}, we expect scenarios with higher traffic demand to be more representative of future applications.
Therefore, we create three outgoing and incoming streams, for each local concentrator, instead of using only a single stream.
To model data exchange between backhaul concentrators, we add incoming and outgoing streams to connect any pair of backhaul concentrators.

Reports of the latency for AMI applications vary from \SI{4}{\milli\second} (more relaxed bounds in residential areas) \cite{Lopez2015} to 12--\SI{20}{\milli\second} \cite{Yan2013}.
Therefore, we use periods (and thus deadlines) of \SI{4}{\milli\second}, \SI{10}{\milli\second}, and \SI{20}{\milli\second}, resulting in a hyper cycle of \SI{20}{\milli\second}.
Bandwidth is reported to be around \SI{500}{\kilo\bit\per\second} for backhaul networks \cite{Gungor2013}.
Hence, we pick frame sizes randomly from 125, 500, and \SI{1500}{\byte}, to achieve an average bandwidth per stream of \SI{500}{\kilo\bit\per\second}.

We investigate offline planning and online updates.
Thus, we schedule the first iteration offline, where most streams area added to the system.
The streams of some randomly chosen local concentrators are withheld to be added in later iterations, simulating a new local concentrator joining the grid.

\paragraph{Evaluation Results}

\begin{figure*}
	\centering
	\includegraphics[trim={6mm, 5mm, 6mm, 5mm}, clip, width=1\linewidth]{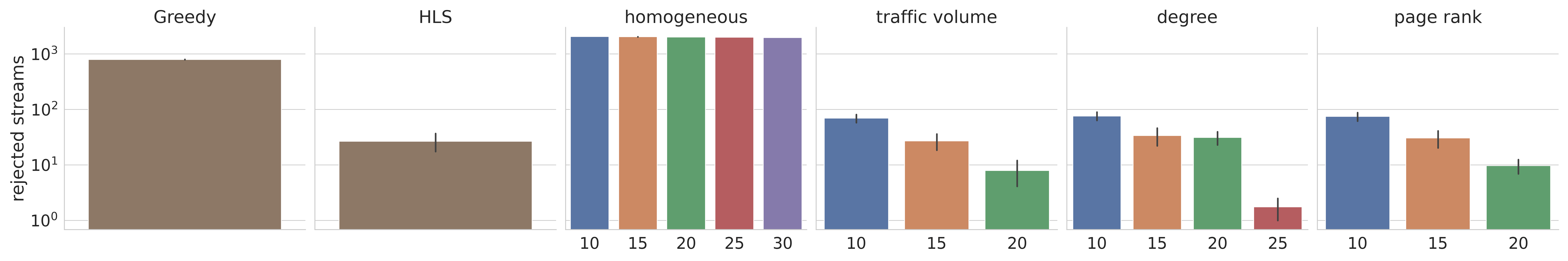}
	\caption{Stream rejection in the advanced metering infrastructure evaluation for varying cps values on the network derived from the IEEE 300-Bus system. Greedy and HLS do not have a CPS parameter.}
	\label{fig:amicatplotrejected-streams}
\end{figure*}

The number of rejected streams for the first (offline) iteration with different enumeration schemes, heterogeneous distribution strategies, and cps values is shown in Figure~\ref{fig:amicatplotrejected-streams}.
Greedy and HSL both fail to schedule all streams. 
Greedy rejects almost 800 streams, HLS ~25.
The deterministic phase enumeration rejects about 2000 streams (out of 2184).
In contrast, the heterogeneous distribution strategies manage to admit all streams with cps=25, except for degree, which admits all streams with cps=30.
For cps=15 all of them perform similar as HLS in terms of rejected streams.
We observe a similar behavior in the online update, where the deterministic phase enumeration rejects most streams, and the other strategies admit most streams, yielding better results as cps increases.
For the higher cps values the heterogeneous distribution strategies reject a single stream in rare border cases.

The total runtime was below \SI{20}{\second} (< \SI{70}{\second} for the deterministic strategy) in the offline planning and finished within less than a second in most online iterations. 
Thereby, the expansion time was responsible for most of the runtime.
Meanwhile, Greedy ran in about \SI{0.4}{\second} and HLS took almost \SI{3}{\minute} for the offline planning.

To summarize, when using our randomized phase enumeration and heterogeneous distribution strategy for the conflict graph expansion, conflict-graph-based scheduling can be used in realistic large-scale systems.
It achieves reasonable offline computation (in the order of seconds) and online update times (in the order of sub-seconds).


\section{Conclusion}
\label{sec:conclusion}

In this paper, we studied the dynamic time-driven traffic planning problem in large-scale time-sensitive networks. 
Conflict-graph-based scheduling is a promising but, until now, not scalable approach, due to the time needed by the conflict graph expansion. 

To this end, we defined and tested a set of conflict graph expansion strategies to make conflict-graph-based scheduling feasible for large-scale systems.
First, we implemented a randomized phase enumeration scheme which greatly improves the scheduling success and conflict graph expansion time compared to a deterministic phase enumeration, by avoiding reoccurring structures in the conflict graph.
Second, we evaluated three heterogeneous distribution strategies, which prioritize some streams over others during the conflict graph expansion.
The \emph{average degree} strategy prioritizes streams based on the average degree of each stream's configuration nodes, while the \emph{traffic volume} strategy assigns more configurations to streams with a low traffic volume.
With these strategies, we obtained particularly good scheduling results within short runtime.

Using these conflict graph expansion strategies, we were able to schedule streams in large-scale networks with 256 bridges and several thousand streams offline within minutes and perform online updates in seconds. 
Further, we implemented the conflict graph expansion with heterogeneous distribution strategies in a realistic energy grid application, and were able to validate their suitability for large-scale real-world network applications.

\label{sec:conclusion:future-work}
So far, we have only tested conflict-graph-based scheduling with unicast communication.
First investigations show that the extension to multicast scheduling is possible, but not trivial. 
Therefore, multicast scheduling is out-of-scope for this publication and will be covered in the future.

\bibliographystyle{ieeetran}
\bibliography{h2s_main.bib}

\end{document}